\documentclass[twocolumn,showpacs,preprintnumbers,amsmath,amssymb]{revtex4}
\usepackage{epsfig}

\begin{document}
\draft


\title{
A ratchet for heat transport between identical reservoirs
}

\author{Souvik Das${}^{1}$\cite{a}, Onuttom Narayan${}^{2,3}$\cite{b} 
and Sriram Ramaswamy${}^{2}$\cite{c}}
\affiliation{${}^1$ St Stephen's College, University Enclave, Delhi 110 007, INDIA\\
${}^2$Centre for Condensed Matter Theory, Department of Physics, 
Indian Institute of Science, Bangalore 560012, INDIA\\
${}^3$ Department of Physics, University of California, Santa Cruz, CA 95064.}

\date{\today}

\begin{abstract}
A one-dimensional periodic array of elastically colliding hard points,
with a non-centrosymmetric unit cell, connected at its two ends to
{\em identical but non-thermal} energy reservoirs, is shown to carry a
sustained unidirectional energy current.
\end{abstract}


\maketitle 
A particle subjected to fluctuating forces far from thermal equilibrium,
in a medium with a vectorial asymmetry, must drift in a direction
determined by the asymmetry, even if the fluctuating forces have zero
mean.  This idea can be seen as a consequence of the ``Curie Principle''
\cite{curieprincip} -- whatever is not ruled out by symmetries is
permitted and therefore obligatory. Reviews of the many realizations of
this concept of a ``Brownian ratchet'' and their relevance to the working
of motor proteins in living cells as well as to novel particle separation
methods can be found in \cite{prostrmp,reimann,linke}. Implementations of the
ratchet idea generally involve randomly forced particles in periodic,
non-centrosymmetric potentials. A nonequilibrium stationary state is
maintained either by having the potential alternate between two states
or by choosing the random forcing not to obey a fluctuation-dissipation
relation with respect to the damping.

Can these methods for particle transport without chemical-potential
gradients be extended to generate {\em energy} currents without {\em
temperature} gradients? We know, of course, that energy cannot flow
spontaneously from one {\em thermal} bath to another with identical
thermodynamic parameters, no matter how asymmetrical the conducting
medium linking the two baths. The requirements of thermal equilibrium and,
hence, of time-reversal invariance, mask the structural asymmetry of the
conductor. We conjecture, therefore, that an energy current should flow
through a suitably asymmetrical medium if the (identical) baths at the
ends were {\em non-thermal}.

Another motivation for the work presented in this paper is the renewed
interest in heat conduction and energy propagation in momentum-conserving
one-dimensional systems \cite{1dheatothers,1dheatus} with a temperature
gradient imposed by connecting the two ends to heat baths at different
temperatures. For a one-dimensional array of point mass particles with
elastic collisions, if the masses are all equal, the system is integrable
and thermal equilibrium is not achieved. Even with unequal masses,
momentum conservation causes the heat conductivity to depend singularly
\cite{1dheatothers} on the length $L$ of the array, as $L^{1/3}$
\cite{1dheatus}. The extension of these results to the nonequilibrium
regime, where a nonintegrable array is connected to non-thermal baths
at the ends, is clearly of fundamental interest.  \begin{figure}
\centerline{\epsfxsize=\columnwidth \epsfbox{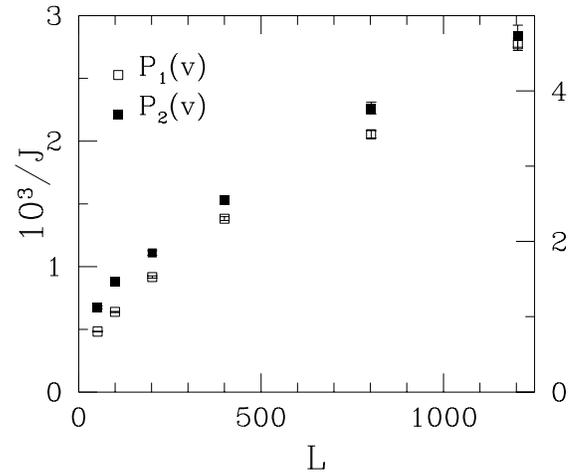}} \caption{
Inverse current flowing from left to right in an ABC array, as a function
of array length $L$. Note that the curve is sublinear, i.e. the current
decays slightly slower than $1/L$. Particles recoil from the baths with
the exponential speed distribution $P_1(v)$ of Eq. (\ref{P1}), or the
heavy tailed distribution $P_2(v)$ of Eq. (\ref{P2}). The $y$-axis for
the $P_1(v)$ and $P_2(v)$ are to the left and right respectively.  }
\label{expon1} \end{figure}

\begin{figure} 
\centerline{\epsfxsize=\columnwidth \epsfbox{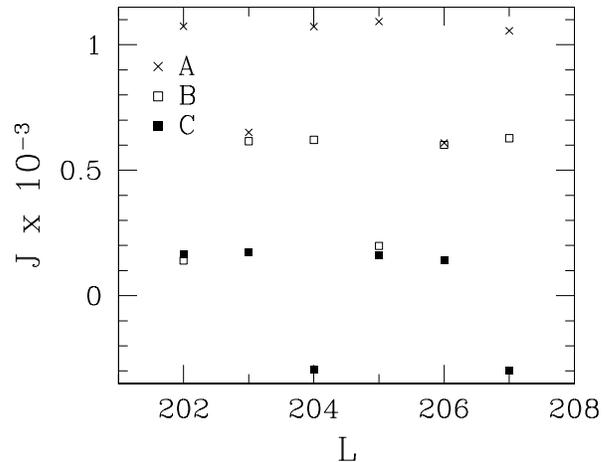}}
\caption{
Current across ABC arrays, for different chain lengths {\em and}
terminating particles. The sets of points marked A, B, and C in the legend
correspond respectively to sequences starting with A, B and C particles.
The velocity distribution at the ends is the exponential $P_1(v)$
of Eq.(\ref{P1}).}
\label{endings}
\end{figure}

\begin{figure}
\centerline{\epsfxsize=\columnwidth \epsfbox{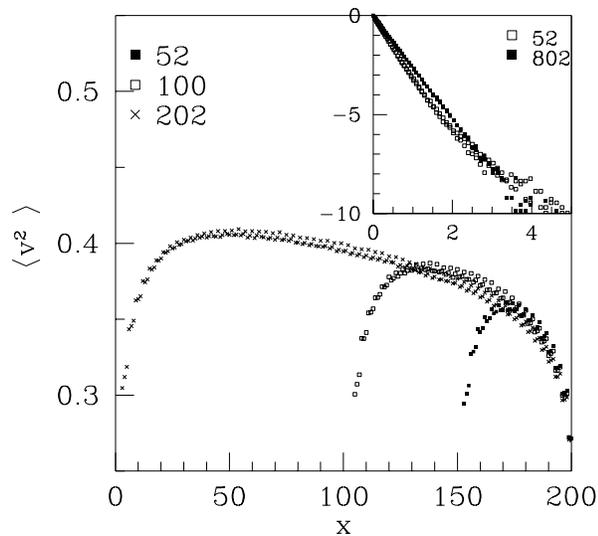}}
\caption{Average kinetic energy of particles across chains of lengths
$L = 52,$ 100, 202. The right hand edges of all the chains are made to
coincide, to show the approximate $L$-independence of the right boundary
region. The length of the left boundary region is also approximately
independent of $L.$ The interior region lengthens as $L$ is increased,
and the kinetic energy varies slowly here. There is a weak cycle-of-three
periodicity to the kinetic energy, whose cause is not clear. The inset
plots the velocity distribution, $\ln[P(v)/P(0)]$ vs $v^2, $ for the
particle beside the center.  There is significant deviation from a
gaussian for $L=52,$ but not for $L=802.$}
\label{gauss}
\end{figure}

\begin{figure}
\centerline{\epsfxsize=\columnwidth \epsfbox{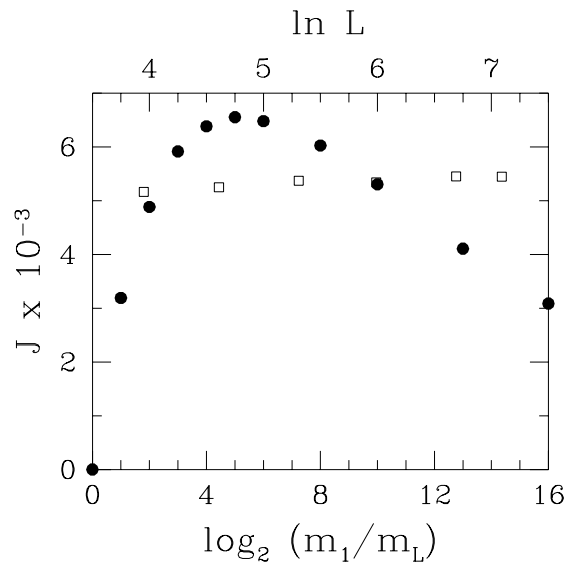}}
\caption{
Current in the `geometric' array, as a function of array length $L$
with an end-to-end mass ratio $m_1/m_L = 5$ (open squares), and as a
function of $\log_2(m_1/m_L)$ for $L=100$ (solid circles).  There is
little variation seen as a function of $L$ despite the logarithmic scale,
in contrast to the pronounced maximum as a function of $m_1/m_L.$ The
error bars are less than the point sizes.  The baths emit particles with
the long-tailed speed distribution $P_2(v)$ of Eq. (\ref{P2}).}
\label{monot1}
\end{figure}

We first present a brief summary of our results and then discuss
our studies in more detail. We study numerically the dynamics of
one-dimensional arrays of particles labelled $i = 1, ... L$ from left
to right, with masses $\{m_i\}$ arranged in a left-right asymmetric
sequence.  The particles are hard points which interact only upon
contact, undergoing elastic collisions.  To the two ends of these arrays
we attach {\em identical, non-thermal} reservoirs of kinetic energy:
when the first or last particle collides with the reservoir beside it,
it recoils with a velocity drawn from a distribution which is not of
the form $P(v)\propto v\exp[-\beta m v^2].$ Regardless of the type
of non-thermal reservoir used and the nature of the asymmetry of the
array, we always find a net macroscopic energy current, confirming our
conjecture above and demonstrating the robust nature of this remarkable
``energy ratchet''. For one class of arrays, the steady-state current $J$
decays slowly with the system length $L$, roughly \cite{anomaly} as $1/L$,
the behavior expected of a conventional heat current in response to a
fixed temperature difference across a length scale $L$. For another,
rather different class of arrays, $J$ is found to be independent of
$L$. In either case, it is clear that the current is not merely an edge
effect present only near the nonequilibrium reservoirs. Systematics
of the dependence on size, sequence, and termination can be found in
Figs. \ref{expon1} to \ref{monot1}.

We now present our study in more detail.  We illustrate our general
point with two types of arrays: (a) the `ABC' array, consisting of three
species of particle A, B, and C with distinct masses $m_A < m_B < m_C$,
arranged in the periodic but non-centrosymmetric sequence ...ABCABCABC...,
and (b) the `geometric' array where $m_{i+1}/m_i = r=$ constant . Since
the dynamics of each particle is purely ballistic between collisions,
we adopt an ``event-driven'' approach: given the present momenta of all
particles, one knows which pairs will next collide and when and, hence,
the post-collision momenta and energies. If the outermost particle at
either end collides with the reservoir, it is re-emitted into the system
with a speed $v$ chosen from a distribution $P(v).$ More precisely, the
rebound speed is first drawn from $P(v)$ and then multiplied by $1/\sqrt{2
m}$ where $m$ is the mass of the particle, so that particles recoiling
from the reservoirs have the same typical kinetic energy regardless of
their mass. We considered two forms for $P(v)$:
\begin{equation}
\label{P1}
P_1(v) = \exp(-v)
\end{equation}
and 
\begin{equation}
\label{P2}
P_2(v) \propto  v/(1 + v^2)^3.
\end{equation}
Both these distributions are easily generated from the standard uniform
random variable over $[0,1).$ The former is somewhat unusual, since
if $P(v)$ is taken to be the distribution for particles leaking out
of a large homogeneous reservoir, the velocity distribution inside
the reservoir must be $p(v)\sim P(v)/v$ which, for (\ref{P1}), is not
normalizable.  However, such an assumption may be too restrictive: the
velocity distribution of particles in a granular flow bumping against
the side walls has under certain circumstances been found \cite{exp-v} to
be $\sim\exp(-v)$, and other non-equilibrium methods of energy injection
at the walls leading to $P_1(v)$ could be conceived of.

We discuss the ABC array first, for which our studies used $m_A = 1, \,
m_B = 2.5, \, m_C = 6.5$, with $P_1(v)$ [Eq. (\ref{P1})], and $m_A = 1.0,
\, m_B = 2.5, \, m_C = 9.0.$ with $P_2(v)$ [Eq. (\ref{P2})].  The mass
values were selected in a rough attempt to maximize the current flowing,
at least for small arrays. In both cases, lengths $L = 52, 100, 202,
400, 802, 1204$ were studied, chosen so that the particle at the end
of the array was always an A.  For both choices of bath distribution,
the steady state has a current $J$ flowing from the left to the right
of the chain and decreasing slowly with $L$. Whether we use $P_1(v)$ or
$P_2(v),$ the decay is no faster than the $1/L$ of ``normal" conduction,
and is clearly not an exponential decay (see Fig. \ref{expon1}).

Although the current decays slowly with chain length, it is sensitive to
the termination of the chains.  To illustrate this, we study the current
$J$ separately for arrays starting with $A$, $B$ and $C$, as a function
of their length. For arrays of short lengths that start with $A$, the
current is non-monotone with a cycle-of-three periodicity. A similar
periodicity is seen when the chain starts with $B$ or with $C.$ There
is significant variation between the three cases.  This is summarized
in Fig. \ref{endings}, where the velocity distribution used was $P_1(v).$

The sensitivity of the current to the termination of the chains might seem
surprising, in view of our earlier conclusion (Fig.~\ref{expon1}) that
the flow of current is not an end effect restricted to short chains. {\it
A priori}, one could have proposed various distinct ways in which the
energy current $J$ might have depended on the system length $L.$ In the
first scenario, repeated collisions in the interior would dominate for
large sizes. The baths at the ends, though nonequilibrium, would merely
act as crude thermostats. It would then be tempting to conclude that
$J\sim\exp(-L/\xi),$ negligible beyond some thermalization length $\xi.$
At the other extreme, the non-thermal nature of the baths could dominate
all other effects and give a current independent of $L.$ Our results for
for the length and termination dependence of the current in ABC arrays
suggests a scenario between these two extremes: interparticle collisions
might lead to a rapid thermalization as one proceeds from the baths
into the interior of the chain, but with {\it different\/} temperatures
near the two ends because of the asymmetric chain and nonequilibrium
baths. If this scenario is correct, the difference between the effective
temperatures near the two ends would depend on the terminations, but with
the terminations specified, the energy current in a large chain would
be due to the effective temperature difference, and would decay with
$L$ as slowly as for thermal conduction. We have measured the average
kinetic energy across the chain for different values of $L.$ As shown in
Fig.~\ref{gauss}, we find a rough separation into two boundary regions
whose size does not vary significantly with $L,$ and an interior region
(whose length increases with $L$) with a slowly varying average kinetic
energy. However, surprisingly, the kinetic energy in the interior is
higher than at the two ends. The velocity distribution near the center
approaches a gaussian as the chain length is increased.

The existence of a non-zero energy current with athermal baths can be
explained qualitatively. If one considers a chain of only two particles,
when the heavy particle recoils from the bath at its end and collides with
the light particle, it typically causes the light particle to rebound and
itself continues to move forward. The energy it was carrying continues
to move forward, although distributed between both particles. But if the
light particle carries some energy from the reservoir at its end to the
heavy particle, most of the energy is reflected back and returned to
the reservoir it came from. Thus we see that energy transport is more
efficient in one direction compared to the other. On the other hand,
the light particle has much higher typical speeds, and therefore impinges
against its reservoir much more often. Thus there is a competition between
the relative inefficiency with which the light particle transmits energy
to the heavy one, and its higher attempt rate. Which one of these wins
depends on the parameters of the system; we have numerically verified
(for baths given by $P_1(v)$) that as the mass ratio is changed,
the current even changes sign. For (identical) thermal baths, the two
competing effects cancel exactly.

We now turn to the second type of array we studied: the geometric
array, where the ratio of the masses of any two subsequent particles is
$m_{i+1}/m_i=r,$ where $r$ is independent of $i.$ The geometric array
was studied using $P_2(v)$ only.  First, a fixed end-to-end mass ratio
$m_L/m_1 = 0.2$ was chosen, and the chain length varied. (The mass ratio
$r$ was therefore closer to unity for longer chains.)  Chain lengths
were 52, 100, 202, 400, 802, 1204. A large steady state current flows
from right to left. As the chain length increases, $J$ asymptotes to a
constant value, {\it independent\/} of $L,$ in striking contrast to what
is seen in the ABC array.  Since $J$ is (roughly) insensitive to $L$ for
fixed $m_L/m_1,$ data was also taken with a fixed chain length of 100,
the mass at the right end $m_L$ equal to 1, and the mass at the left
end varied. Not surprisingly, this yields a current that is zero when
$m_1 = 1$, peaks as $m_1$ is increased, and then decays again as $m_1$
is increased further. The peak is approximately at $m_1= 100$.

These results can be justified by recalling that for a chain where all
the particles have equal mass, energy propagation is ballistic. With {\em
thermal} baths at different temperatures at the two ends, such a chain
carries an $L$-independent current.  (There would be no current with
identical baths at both ends for such a chain.) Thus for a monotonic
chain, where the masses vary only gradually, one should measure
the length in terms of the ``scattering length" over which transport
deviates appreciably from the ballistic, i.e. the length over which the
mass changes by some reasonable factor. If $m_L/m_1$ is held fixed, the
``effective length" is then independent of $L$. A chain with a very small
effective length would look symmetric, and therefore not carry energy
between identical reservoirs, while the current in a chain with a very
large effective length would diminish because of repeated scattering.
Although qualitative, this argument would suggest that the scattering
length should roughly correspond to $m_1/m_L = 100,$ for which $J$
is maximum.

The two sets of observations on the geometric chain, taken together,
lead to a surprising conclusion: if one takes a (long) monotonic chain
with $m_1 = 10$ and $m_L = 1$, and another chain with the same ratio
between successive masses, with $m_1 = 100$ and $m_L = 10$, the current
in the first chain should be sqrt(10) times the current in the second,
but when the two chains are connected in series (in the correct order),
the current will be greater than that in either of its two components.

Several natural extensions of the work presented here suggest
themselves. For example, consider the case where the baths at the ends
are in identical steady states which are characterized by a parameter
$\lambda$ such that $\lambda =0$ corresponds to thermal equilibrium. How
does the ratchet energy current depend on $\lambda$?  Another intriguing
issue is the effect of inelastic collisions, which are crucial if this
idea is to be applied to granular matter.  An analytical understanding
of our results would clearly be of great interest. Most exciting would
be an experimental realization, perhaps a suitably nano-engineered
wire in the form of a one-dimensional array with an asymmetric unit
cell, with the ends driven by cells containing identical, exothermic
chemical reactions, or perhaps a row of ball-bearings with particle
masses varying as in our ABC or geometric arrays, driven at the ends
by gas-fluidization. The analysis of the latter would have to include
inelasticity, but it should nonetheless display an energy current.

To summarize: our work proposes a new mode of energy transport, in
rough analogy to Brownian ratchets but with features quite distinct from
those systems.  The heat transport here requires no manipulation of the
bulk of the one-dimensional medium, only a structural asymmetry. Only
the two ends of the ``conductor'' are held in (identical) non-thermal
steady states. Neither a time-dependent potential nor noise that violates
detailed balance is required in the bulk. Provided energy is supplied at
both ends in the same nonequilibrium manner, the classical dynamics of
the intervening medium extracts energy preferentially from one
reservoir and sends it to the other. No violation of the second law of
thermodynamics occurs: we have simply specified the properties of the
nonequilibrium reservoirs without saying how they are to be maintained
that way.

SD and SR acknowledge partial support, respectively, from the Kishor
Vaigyanik Protsahan Yojana, India, and the Institute for Theoretical
Physics, UC Santa Barbara (National Science Foundation Grant
No. PHY99-07949).


\begin{references}

\bibitem[a]{a}souvik@vsnl.com
\bibitem[b]{b}narayan@wagner.ucsc.edu
\bibitem[c]{c}sriram@physics.iisc.ernet.in

\bibitem{curieprincip} P. Curie, J. Phys. (Paris) {\bf 3}$^o$ S\'{e}rie 
(th\'{e}orique et appliqu\'{e}) t. III, 393 (1894).

\bibitem{prostrmp} F. J\"ulicher, A. Ajdari, J. Prost, Rev. Mod. Phys. {\bf 69},
1269 (1997) and references therein.

\bibitem{reimann}  P. Reimann, Phys. Rep. {\bf 361}, 57 (2002); cond-mat/0010237
and references therein.

\bibitem{linke} Applied Physics A (Materials Science \& Processing)
{\bf 75}, 167-352 (Special issue on ``Ratchets and Brownian motors'';  
Guest Editor: H. Linke). 

\bibitem{1dheatothers} S. Lepri, R. Livi and A. Politi, Phys. Rev. Lett. 
{\bf 78}, 1896 (1997); T. Hatano, Phys. Rev. E {\bf 59}, R1 (1999); 
S. Lepri, R. Livi and A. Politi, Europhys. Lett. {\bf 43}, 
271 (1998).
T. Prosen and D.K. Campbell, Phys. Rev. Lett. {\bf 84},
2857 (2000); 
A. Dhar, Phys. Rev. Lett. {\bf 86}, 3554 (2001); 
{\it ibid.} {\bf 86}, 5882 (2001).
A.V. Savin, G.P. Tsironis, A.V. Zolotaryuk, Phys. Rev. Lett. 
{\bf 88}, 154301 (2002); G. Casati and T. Prosen, cond-mat/0203331; 
P. Grassberger, W. Nadler and L. Yang, nlin.CD/0203019.

\bibitem{1dheatus} O. Narayan and S. Ramaswamy, cond-mat/0205295 

\bibitem{anomaly} Advection by hydrodynamic velocity fluctuations should 
lead, as in \cite{1dheatothers,1dheatus}, to $J$ decaying slower than $1/L$. 

\bibitem{exp-v} E. Longhi, N. Easwar and N. Menon, cond-mat/0203379.

\bibitem{foot1} We have also measured the kinetic energy
and velocity distribution across the chains, and find that for fixed
$m_L/m_1$ they depend only on the fractional distance $x/L$ down the
chain. Unlike for ABC chains, the velocity distribution near the
center does not approach a gaussian for any $L$ and $m_L/m_1$ that
we have studied.


\end{references}
\end{document}